\documentclass[final,3p,times]{elsarticle}

\usepackage{graphicx}
\usepackage{epstopdf}

\usepackage{amssymb}

\usepackage{amsfonts}
\usepackage{amsmath}

\journal{Optics Communications}

\begin{document}

\begin{frontmatter}



\title{Bragg gap solitons in $\mathcal{PT}$ symmetric lattices with competing nonlinearity}

\author[l1,l3]{Shuang Liu}
\author[l1]{Caiwen Ma}
\author[l2]{Yiqi Zhang\corref{c1}}
\cortext[c1]{Corresponding author.}
\ead{zhang-yiqi@163.com}
\author[l4]{Keqing Lu}

\address[l1]{Xi'an Institute of Optics and Precision Mechanics of Chinese Academy of Sciences, Xi'an 710119, China}
\address[l2]{Department of Electronic Science and Technology, Xi'an Jiaotong University, Xi'an 710049, China}
\address[l3]{Graduate University of Chinese Academy of Sciences, Beijing 100049, China}
\address[l4]{School of Information and Communication Engineering, Tianjin Polytechnic University, Tianjin 300160, China}

\begin{abstract}
The effect of competing nonlinearity on beam dynamics in parity-time $(\mathcal{PT})$ symmetric potentials is investigated.
By using numerical methods,
the existence of gap solitons is demonstrated in the first Bragg gap of optical $\mathcal{PT}$ symmetric lattices
with competing nonlinearity.
Meanwhile, the stability of such solitons is analyzed through introducing a small perturbation to the solitary solutions.
The abrupt annihilation of the solitons during propagation demonstrates the Bragg gap solitons in $\mathcal{PT}$ symmetric potentials
are not stable.
In comparison with the on-site gap solitons, the off-site gap solitons exhibit more robust properties during propagation.
\end{abstract}


\end{frontmatter}


\section{Introduction}
In quantum mechanics, all physical observables correspond to the eigenvalues of operators demand that
the eigenvalues should be real and thus must be Hermitian.
Yet in recent years a new concept has been proposed in an attempt to extend the framework of quantum mechanics into the complex domain.
It is found that it is in fact possible even for non-Hermitian Hamiltonians to exhibit entirely real eigenvalue spectra as long as
they respect parity-time ($\mathcal{PT}$) symmetry\cite{bender_prl_1998, bender_prl_2002, bender_prl_2007}.
$\mathcal{PT}$ symmetry means that the eigenfunctions of a Hamiltonian are at the same time the eigenfunctions of
the $\hat{P}\hat{T}$ operator,
that is $H \hat{P}\hat{T} = \hat{P}\hat{T} H$.
Generally, the action of the parity operator $\hat{P}$ is defined by the
relations $\hat{p} \rightarrow -\hat{p},~\hat{x} \rightarrow -\hat{x}$,
whereas that of the time operator $\hat{T}$ by $\hat{p} \rightarrow -\hat{p},~\hat{x} \rightarrow \hat{x},~i \rightarrow -i$,
where $\hat{p}$ and $\hat{x}$ represent the momentum and position operators, respectively.
From this point of view, it is easy to find that a $\mathcal{PT}$-symmetric Hamiltonian requires
$\hat{p}^2/2+V^*(-\hat{x}) = \hat{p}^2/2+V(\hat{x})$,
which indicates that the real part of the complex potential should be an even function of position and the imaginary part should be an odd one.
It is noteworthy to stress that this condition is just necessary but not sufficient.

To date, spatial solitons (localized bound states that can maintain their shapes during propagation
in a bulk media\cite{zhang_cpl_2009,zhang_cpb_2009} or waveguide\cite{zhang_oe_2010})
in periodic optical lattices with $\mathcal{PT}$ symmetry are quite
involved\cite{musslimani_prl_2008, makris_prl_2008, ruter_np_2010, zhou_ol_2010, wang_oe_2011}.
In this article, we investigate the gap solitons in $\mathcal{PT}$ symmetric lattices with competing nonlinearity for the first time.
The competing nonlinearity adopted in our model is the so-called cubic-quintic (CQ) nonlinearity.
CQ nonlinearity contains two parts (proportionally to the beam intensity and the intensity square, respectively) with different signs,
that the nonlinearity induced by the beam is greatly affected by the intensity, i.e.,
with different functionalities with respect to the intensity,
the nonlinearity may change from self-focusing to self-defocusing, or from self-defocusing to self-focusing
along the transverse profile.
Even though it is reported that stable (gap) solitons are demonstrated in $\mathcal{PT}$ symmetric lattices with Kerr nonlinearity\cite{musslimani_prl_2008},
in the models combined periodic potentials\cite{merhasin_pre_2005} or the Bragg coupled-mode structure\cite{atai_pla_2001} and the CQ nonlinearity,
and in a complex Ginzburg-Landau system\cite{sakaguchi_pre_2008},
the results we obtained here are quite different.
The article is organized as follows: In Section 2, we briefly introduce the general model equation
for beam propagation in $\mathcal{PT}$ symmetric lattices with competing nonlinearity.
In Section 3, firstly, we discuss the Bloch band structures of the complex potential and give the diagrams of the structures.
And secondly, we investigate the on-site solitary solutions as well as the off-site solitary solutions in the band gap.
Last but not least, in order to do stability analysis of the solitary solutions,
we use beam propagation method to investigate the propagation properties of the solitary solutions with small perturbations.
In Section 4 we conclude the article.

\begin{figure}[htbp]
\centering
  \includegraphics[width=0.3\textwidth]{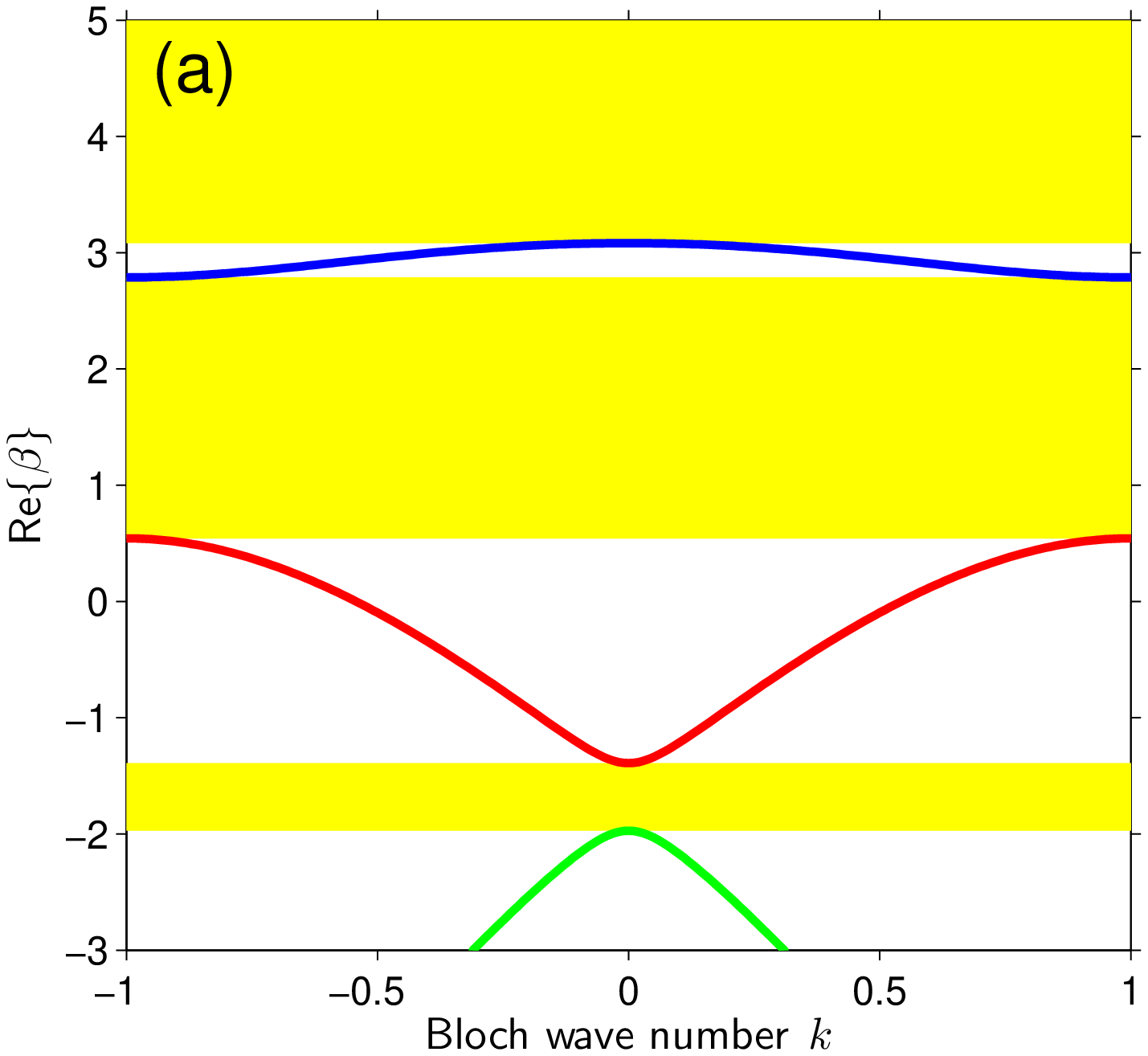}
  \includegraphics[width=0.3\textwidth]{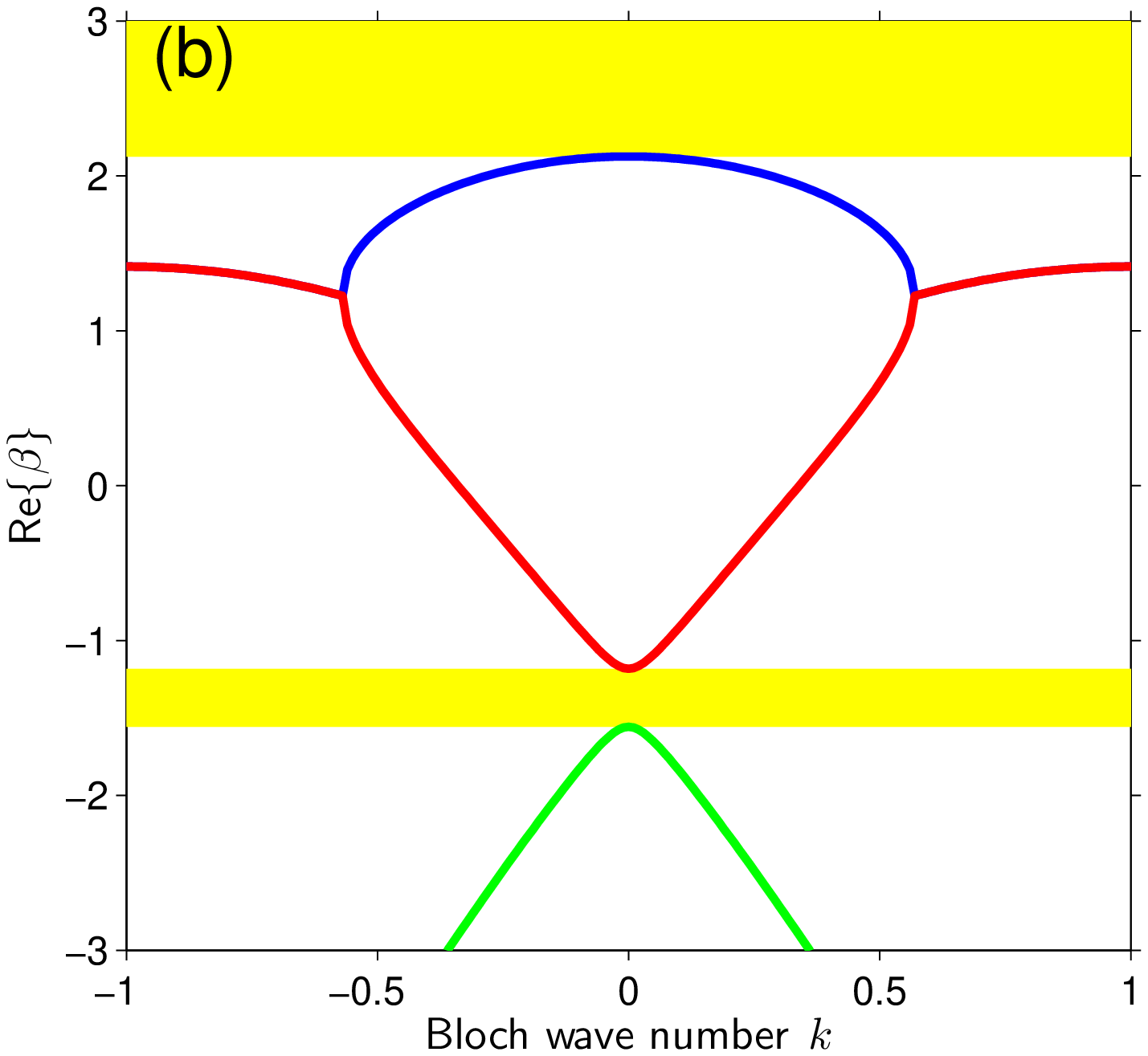}\\
  \includegraphics[width=0.3\textwidth]{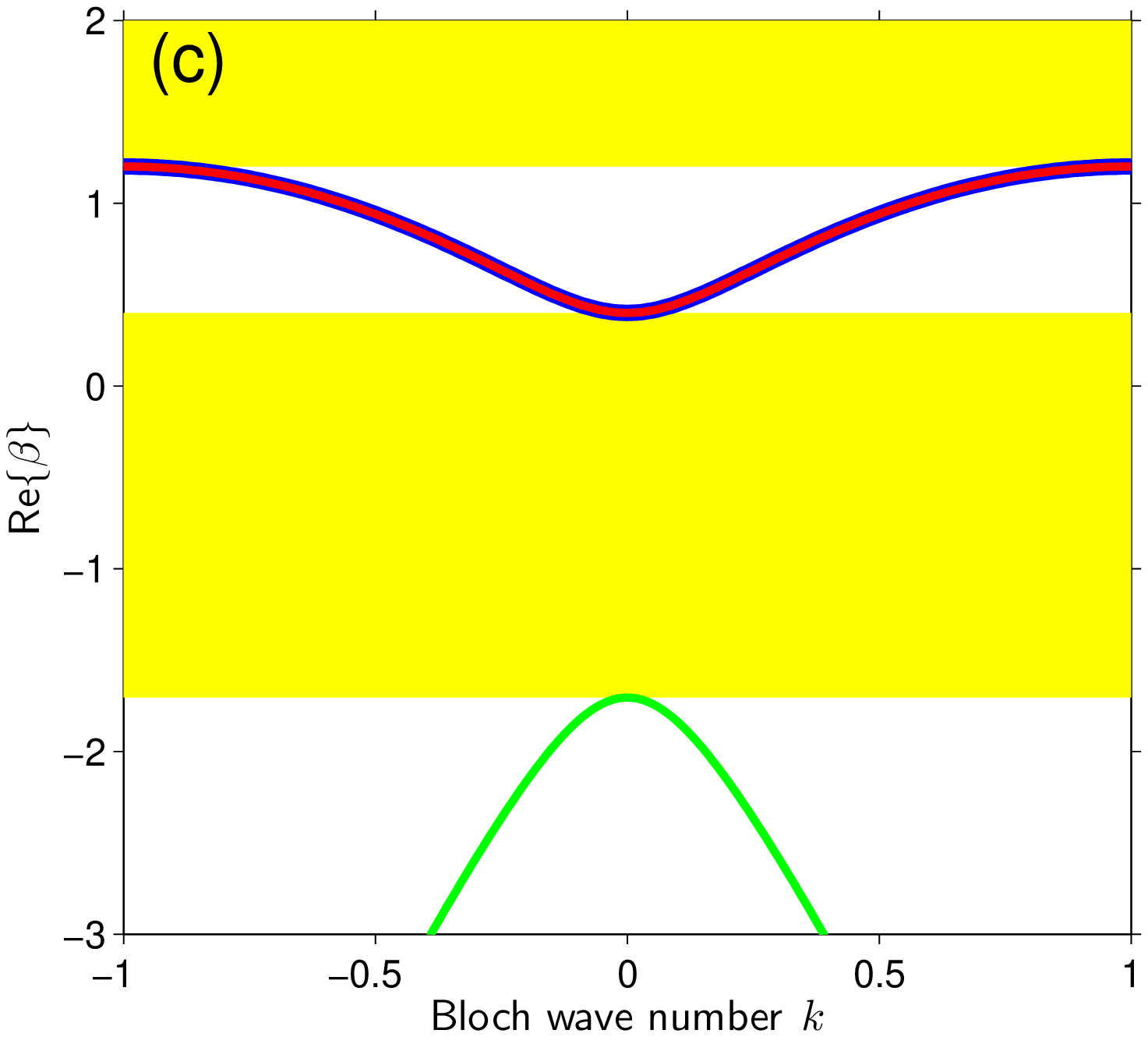}
  \includegraphics[width=0.3\textwidth]{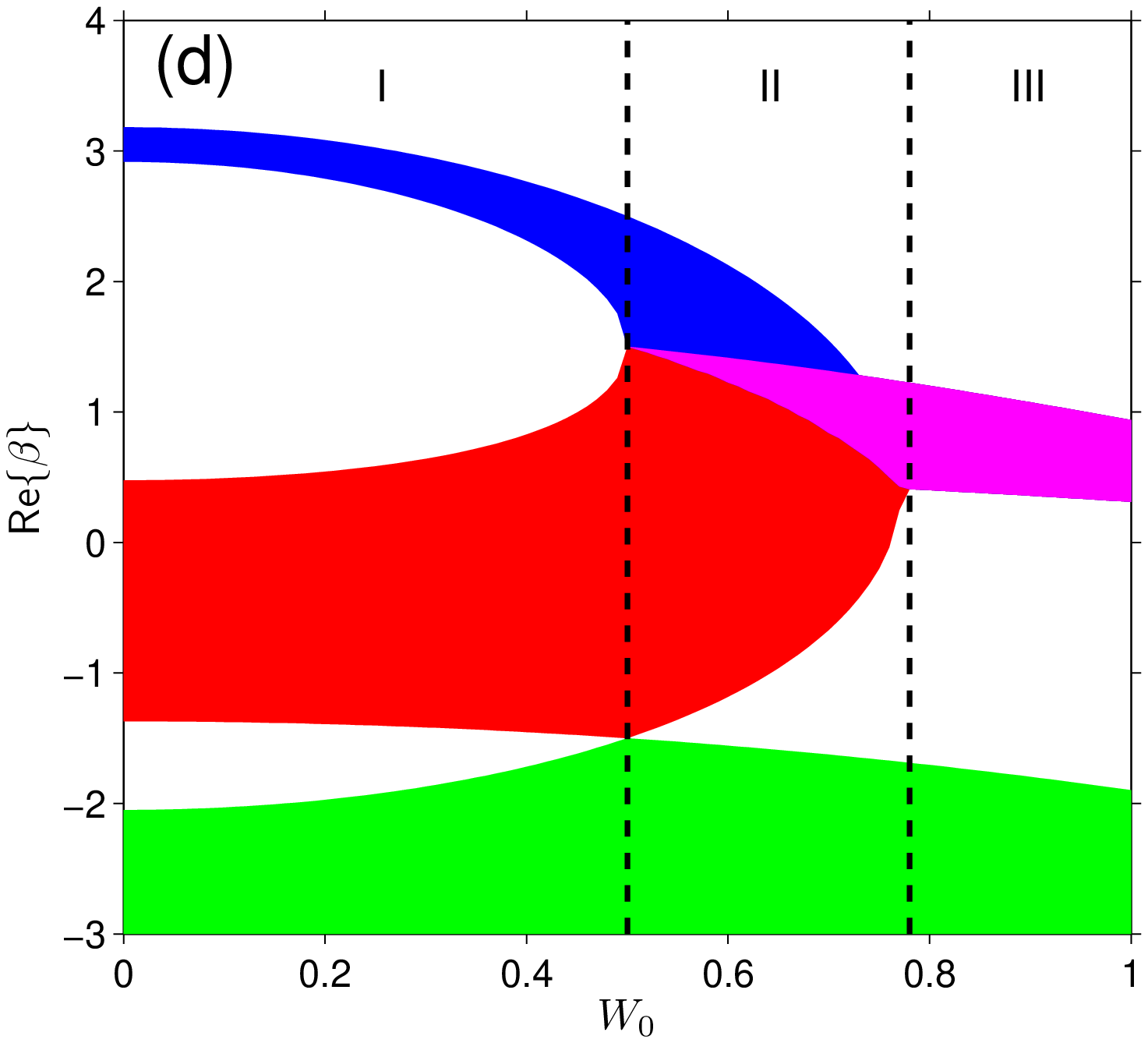}
  \caption{(a)-(c) Band structures corresponding to $W_0=0.2,~0.6$ and $0.8$, respectively.
  The curves are the Bloch bands and the yellow regions are the band gaps.
  (d) real parts of the eigenvalues as a function of $W_0$.
  In part I, all the eigenvalues are purely real;
  in part II, the eigenvalues in the cone-like regions are real and those in other places are complex;
  in part III, the eigenvalues are completely complex.
  The colored regions are the Bloch bands and the blank regions are the band gaps.  }
  \label{bands_diagram}
\end{figure}

\section{Theoretical model}
We begin our analysis by considering optical wave propagation in a competing nonlinear $\mathcal{PT}$ symmetric potential,
which is governed by the traditional normalized nonlinear Schr\"odinger equation as used in previous literatures\cite{musslimani_prl_2008}
\begin{equation}\label{nlse}
    i\frac{\partial \psi}{\partial z} + \frac{\partial^2 \psi}{\partial x^2} + A_0 \left[ V(x) + i W(x) \right] \psi + \left( I - I^2 \right) \psi= 0,
\end{equation}
where $I=|\psi|^2$ is the beam intensity, $I-I^2$ correspond to a competing CQ
optical nonlinearity\cite{mihalache_pre_2000},
$A_0 [V(x) + i W(x)]$ is the so called complex potential,
and in our simulation we take $A_0=5$.
In spatial domain, the transverse coordinate $x$ and the longitudinal coordinate $z$ are scaled to the input beam width $x_0$ and
the diffraction length $L_\textrm{diff}=n_0k_0x_0^2$, respectively, where $n_0$ is the background refractive index and $k_0=2\pi/\lambda_0$.
According to the necessary condition for a $\mathcal{PT}$ symmetric potential mentioned above,
the real and the imaginary parts of the complex potential should satisfy the relations $V(-x)=V(x),~W(-x)=-W(x)$, respectively.
Similar to the previous literatures\cite{musslimani_prl_2008,makris_prl_2008,zhou_ol_2010,wang_oe_2011}, we consider a relatively simple case:
\begin{equation}\label{potential}
    V(x) = \cos^2(x), \quad W(x) = W_0 \sin(2x),
\end{equation}
where $W_0$ determines the amplitude of the imaginary part
and $\pi$ is the period of the potential.
Generally, the solution to Eq.(\ref{nlse}) has the form $\psi(x) = \phi(x) e^{i\beta z}$ where $\phi(x)$ is the nonlinear eigenmode
and $\beta$ is the corresponding real propagation constant.
Plug the solution into Eq.(\ref{nlse}), we obtain
\begin{equation}\label{soliton}
    \frac{d^2 \phi}{d x^2} + A_0 [V(x) + i W(x)] \phi + |\phi|^2 \phi - |\phi|^4 \phi= \beta \phi.
\end{equation}
In light of the fact that $\phi$ is a complex localized wavefunction, $\phi$ can be written as $\phi = \phi_\mathrm{R} + i \phi_\mathrm{I}$,
where $\phi_\mathrm{R}$ and $\phi_\mathrm{I}$ represent the real part and imaginary part, respectively.
Substituting $\phi_\mathrm{R} + i \phi_\mathrm{I}$ for $\phi$ in Eq.(\ref{soliton}), we end up with
\begin{equation}\label{relaxation}
\begin{split}
\frac{d^2 \phi_\mathrm{R}}{d x^2} + A_0 \left( V\phi_\mathrm{R} - W\phi_\mathrm{I} \right) + \left[ \left( \phi_\mathrm{R}^2 + \phi_\mathrm{I}^2 \right) - \left( \phi_\mathrm{R}^2 + \phi_\mathrm{I}^2 \right)^2 \right] \phi_\mathrm{R} =~& \beta\phi_\mathrm{R}, \\
\frac{d^2 \phi_\mathrm{I}}{d x^2} + A_0 \left( W\phi_\mathrm{R} + V\phi_\mathrm{I} \right) + \left[ \left( \phi_\mathrm{R}^2 + \phi_\mathrm{I}^2 \right) - \left( \phi_\mathrm{R}^2 + \phi_\mathrm{I}^2 \right)^2 \right] \phi_\mathrm{I} =~& \beta\phi_\mathrm{I}.
\end{split}
\end{equation}

\begin{figure}[htbp]
\centering
  \includegraphics[width=0.3\textwidth]{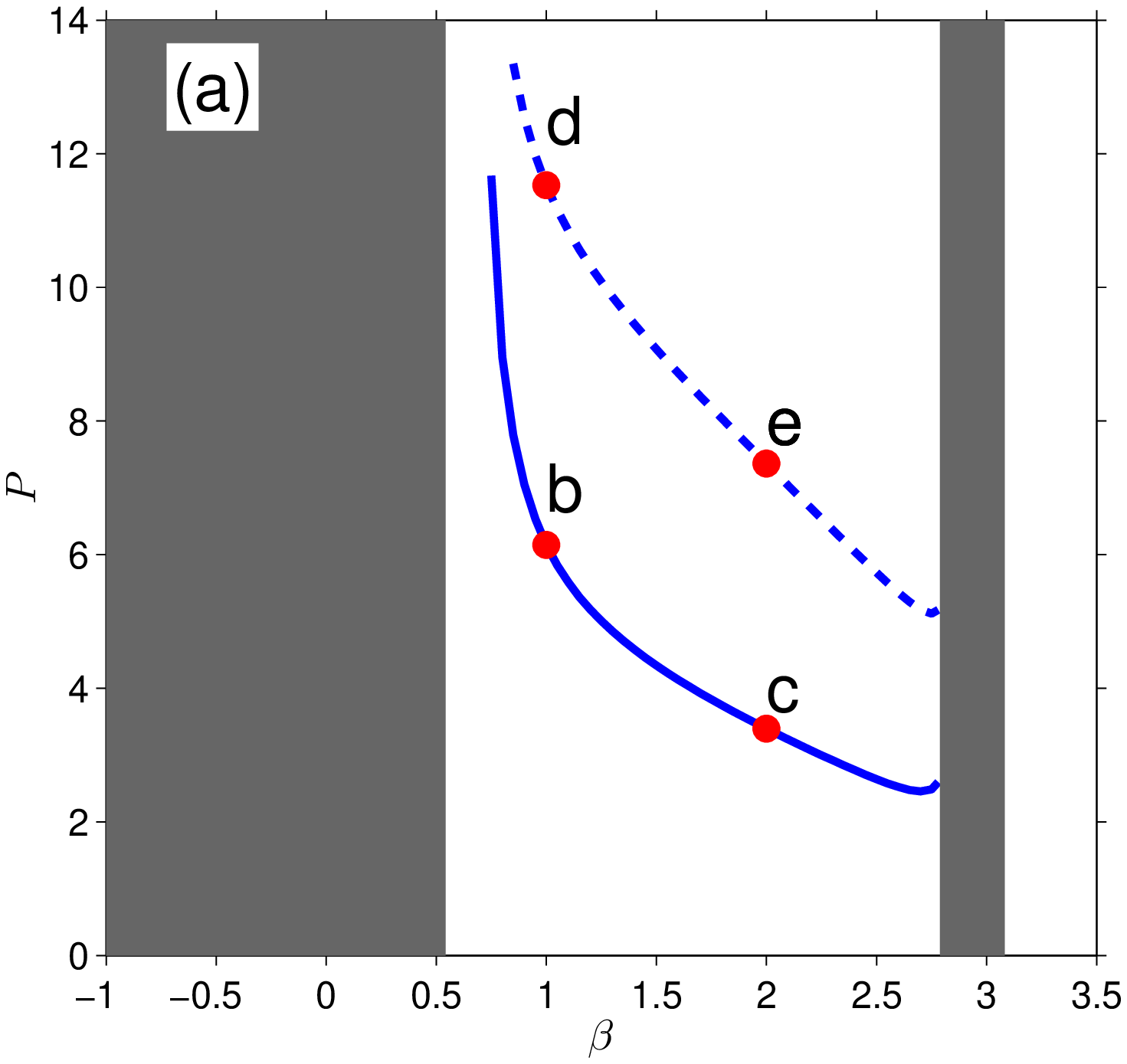} \\
  \includegraphics[width=0.3\textwidth]{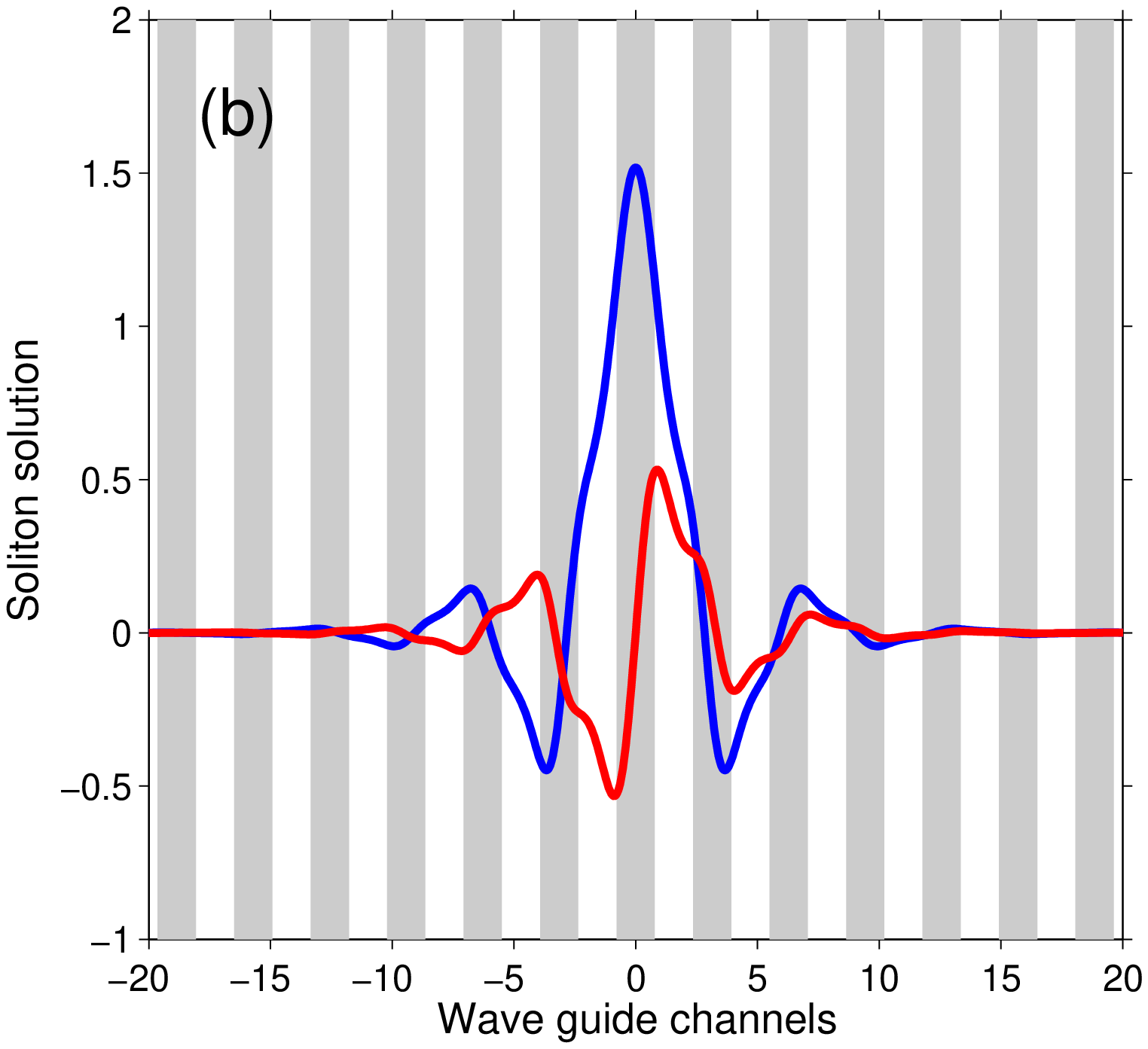}
  \includegraphics[width=0.3\textwidth]{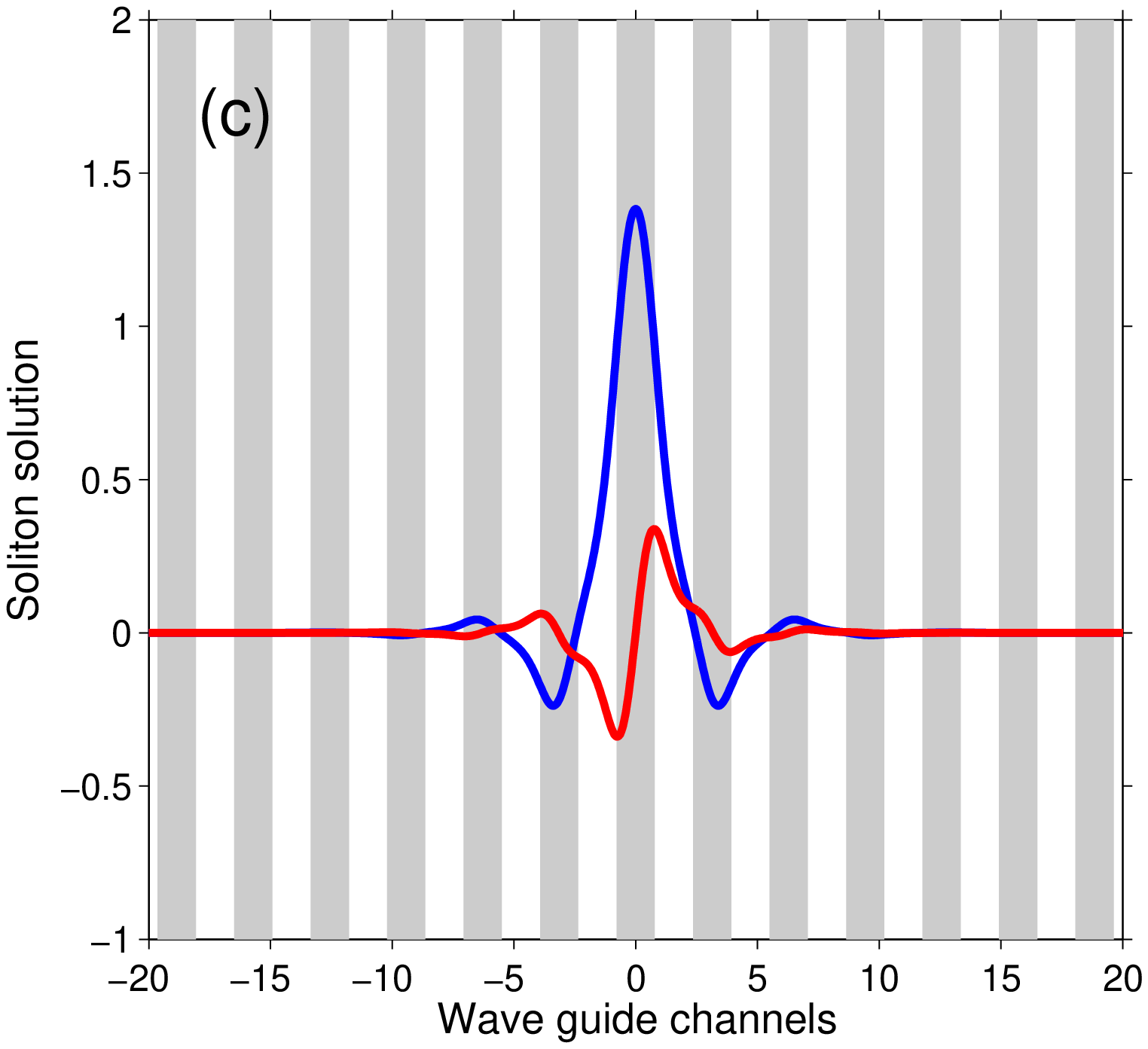} \\
  \includegraphics[width=0.3\textwidth]{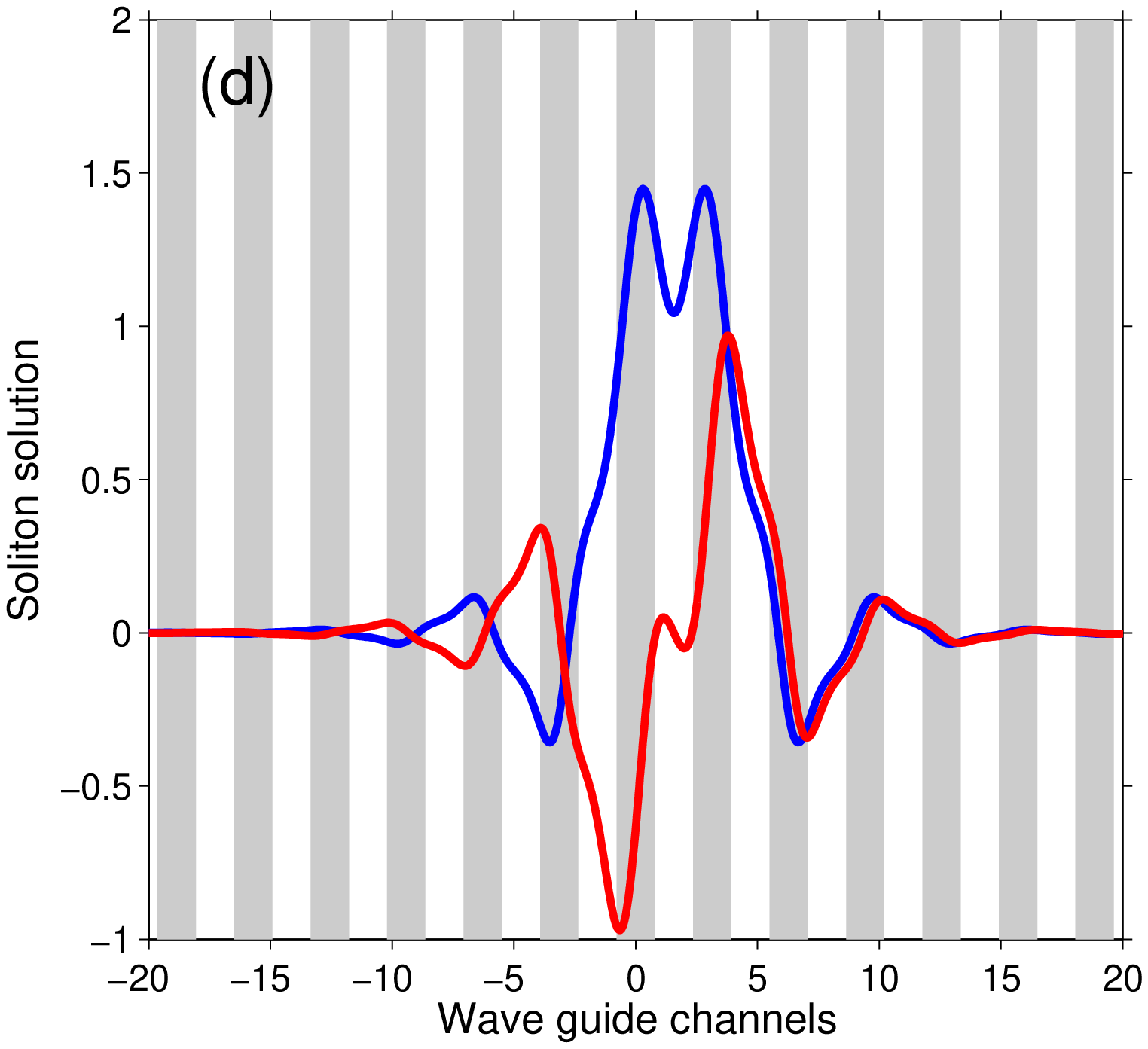}
  \includegraphics[width=0.3\textwidth]{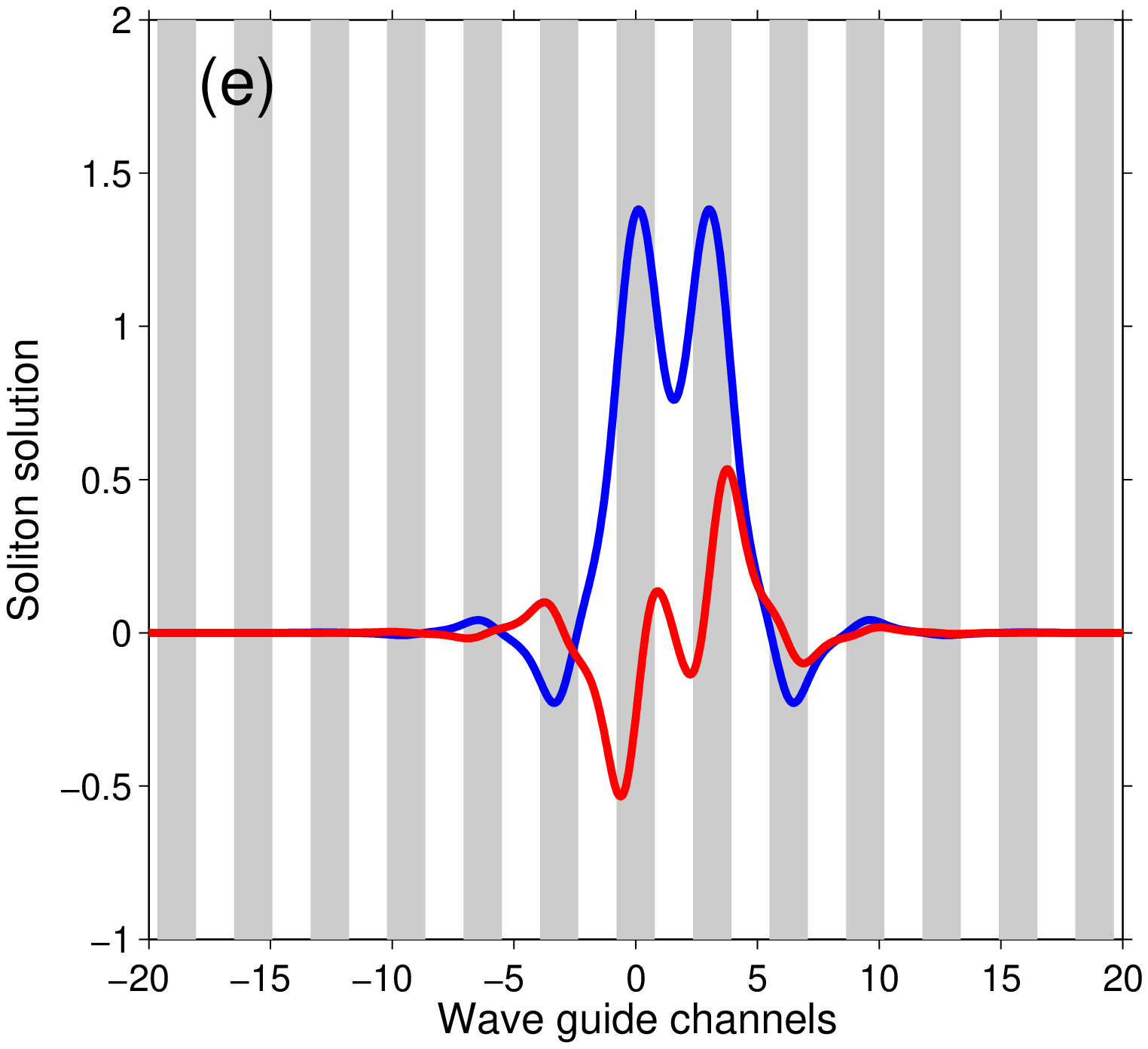}
  \caption{(a) Power curves of solitons (solid curve for the on-site solitons and dashed curve for the off-site ones) in the Bloch gap.
           (b)-(c) $\mathcal{PT}$ on-site soliton field profiles for $\beta=1$ and $\beta=2$, respectively.
           (d)-(e) $\mathcal{PT}$ off-site soliton field profiles for $\beta=1$ and $\beta=2$, respectively.
           The real parts are plotted in blue curves and imaginary parts in red curves.}
  \label{solitons}
\end{figure}

\section{Bragg gap solitons}
Before solving for the localized solutions,
we first analyze the linear properties of such a periodic complex potential by omitting the nonlinear term in Eq.(\ref{soliton}).
As the Floquet-Bloch theorem demonstrated, the eigenfunctions can be written in the following way
\begin{equation}\label{fbtheorem}
    \phi(x) = w_k(x)\exp(ikx), \quad w_k(x) = w_k(x+\pi),
\end{equation}
where $k$ is the Bloch wave number. In this article we just consider the case that $k$ lies in the region $-1 \leq k \leq 1$,
i.e., the first Brillouin zone.
Thus, we can calculate the Bloch band structures corresponding to the complex potential
by using the plane wave expansion method (PWE)\cite{yang_book_2010}.
As pointed in the previous literatures\cite{musslimani_prl_2008,makris_prl_2008,zhou_ol_2010},
there is a critical value $W_0^\textrm{th}=0.5$,
below which the eigenvalues are purely real.
Above the threshold the eigenvalues are partially complex.
Further increasing $W_0$ to a certain value (the value is changing with $A_0$, and for $A_0=5$ the value is about $0.78$),
bands will overlap each other, that indicates all the eigenvalues are complex.
In Fig.\ref{bands_diagram}(a)-(c), we exhibit the three typical band structures for $W_0=0.2,~0.6$, and $0.8$.
The yellow regions are the band gaps.
In Fig.\ref{bands_diagram}(d), the real parts of the eigenvalues changing with $W_0$ is depicted by the colored regions.
The blue, red and green areas are the first, second and third Bloch band regions, respectively.
The pink area represents the region where the first and second bands overlap each other.
According to the critical points of $W_0$,
the bifurcation in Fig.\ref{bands_diagram}(d) is divided into three parts labeled I ($0 \leq W_0 \leq 0.5$),
II ($0.5 \leq W_0 \leq 0.78$) and III ($0.78 \leq W_0 \leq 1$) by two dashed lines.

Similar to the localized solutions in real potentials, in complex potentials the localized solutions can be also divided into two
categories: the on-site lattice solitons and the off-site lattice solitons\cite{lederer_pr_2008}.
And for $W_0 < 0.5$, we solve the coupled Eqs.(\ref{relaxation}) for the family of on-site
as well as off-site localized solutions with real eigenvalues located in the Bragg gap by using the relaxation method.
In Fig.\ref{solitons}(a), we exhibit the power of the on-site (solid curve) and off-site (dashed curve)
solitary solutions in Bragg gap $(0.54<\beta<2.78)$ for $W_0=0.2$.
From Fig.\ref{solitons}(a), firstly we can see that the power of the off-site solitons is bigger than that of the on-site ones,
and secondly we can see that the power for both two type solitons has a minimum value
at $\beta$ very close to the right boundary of the band gap.
Corresponding to the red dots in Fig.\ref{solitons}(a), we display the localized solitary solutions in Fig.\ref{solitons}(b)-(e).
Comparing Fig.\ref{solitons}(b) with Fig.\ref{solitons}(c) and Fig.\ref{solitons}(d) with Fig.\ref{solitons}(e),
and considering the power curves shown in Fig.\ref{solitons}(a),
we can conclude that
(i) the more localized the solitary solution is, the lower the corresponding power is;
(ii) very close to the boundaries of the Bragg gap, it is really hard to observe a solitary solution,
for the localization becomes worse and worse.

\begin{figure}[htbp]
\centering
  \includegraphics[width=0.3\textwidth]{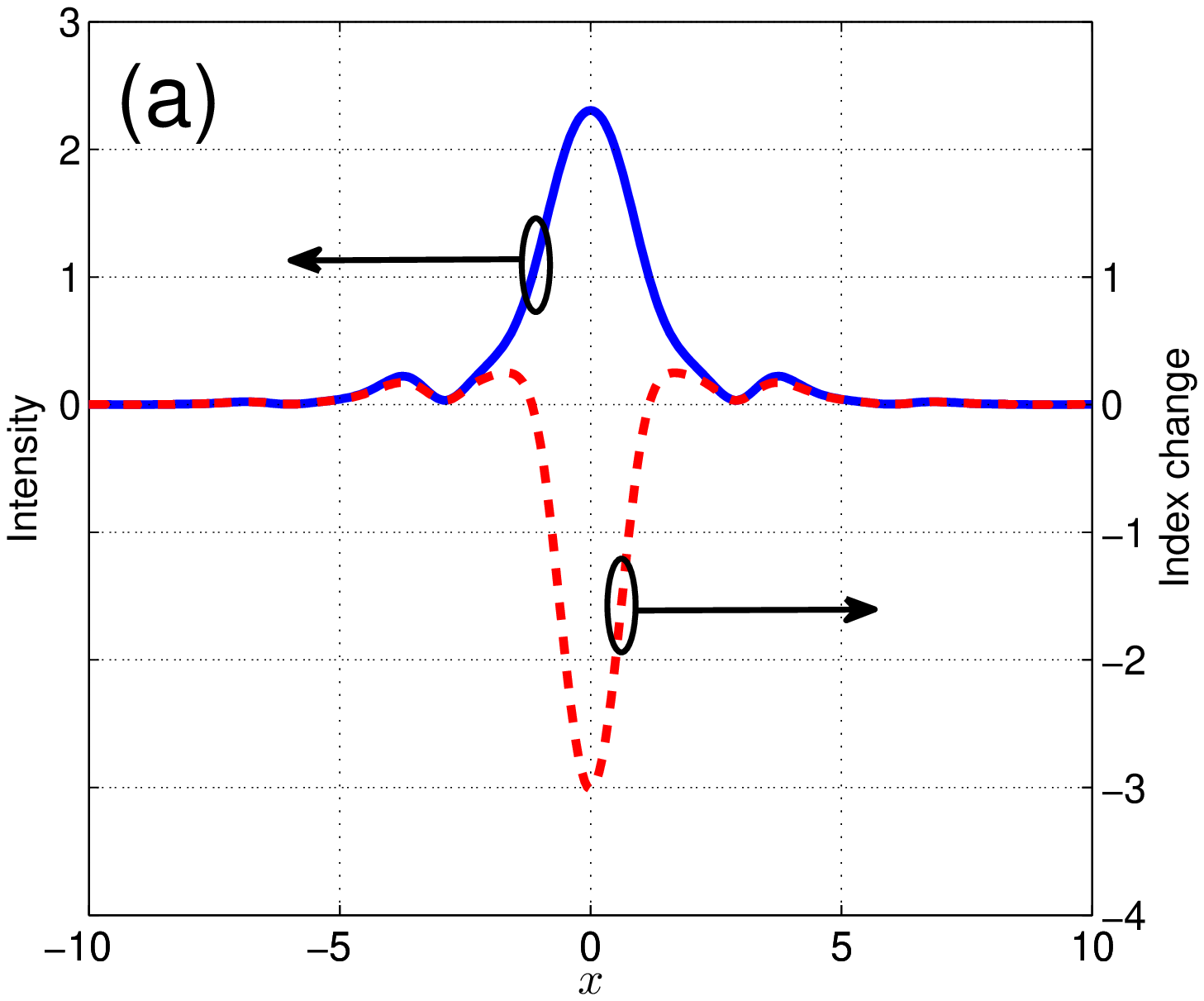}
  \includegraphics[width=0.3\textwidth]{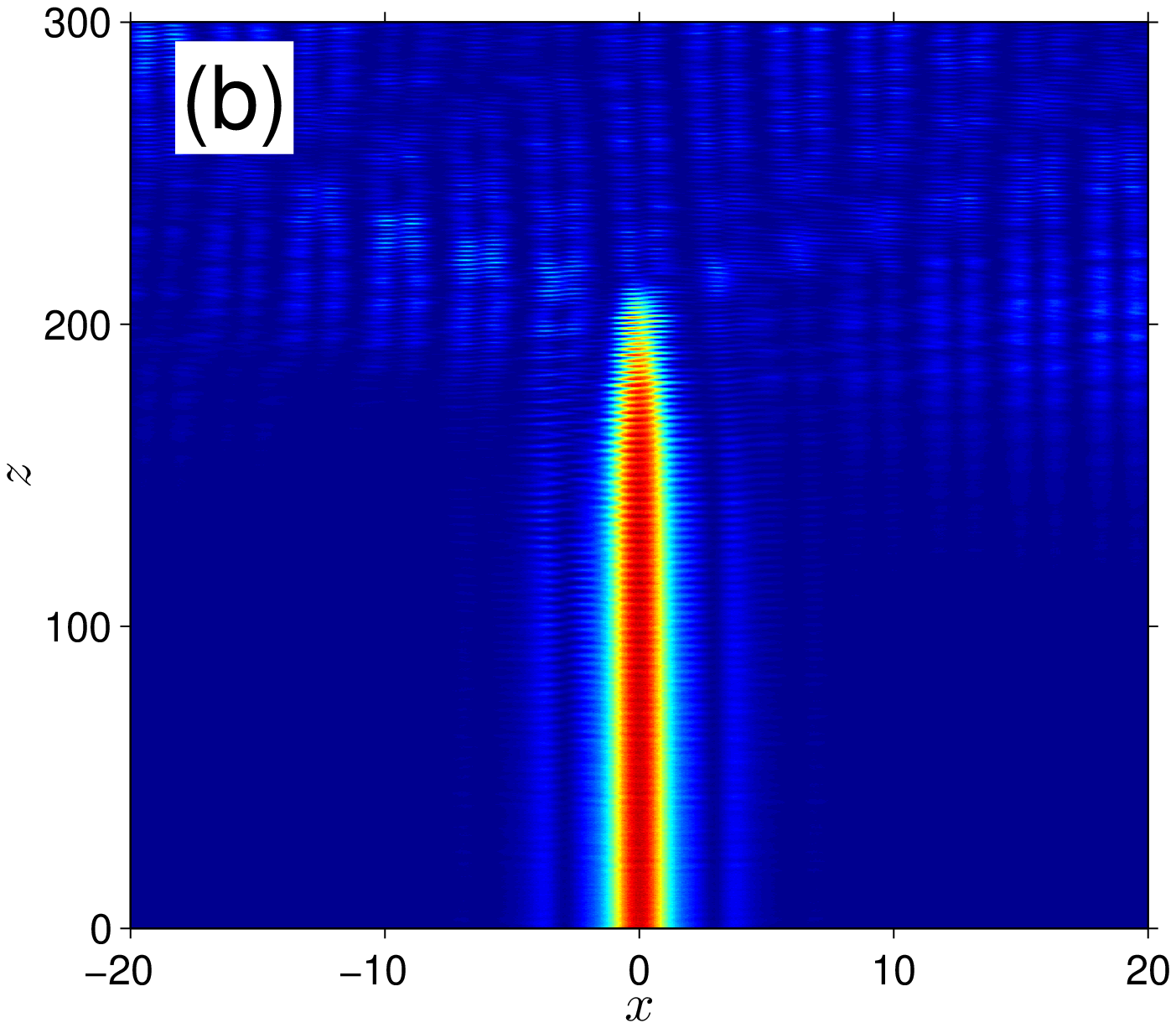} \\
  \includegraphics[width=0.3\textwidth]{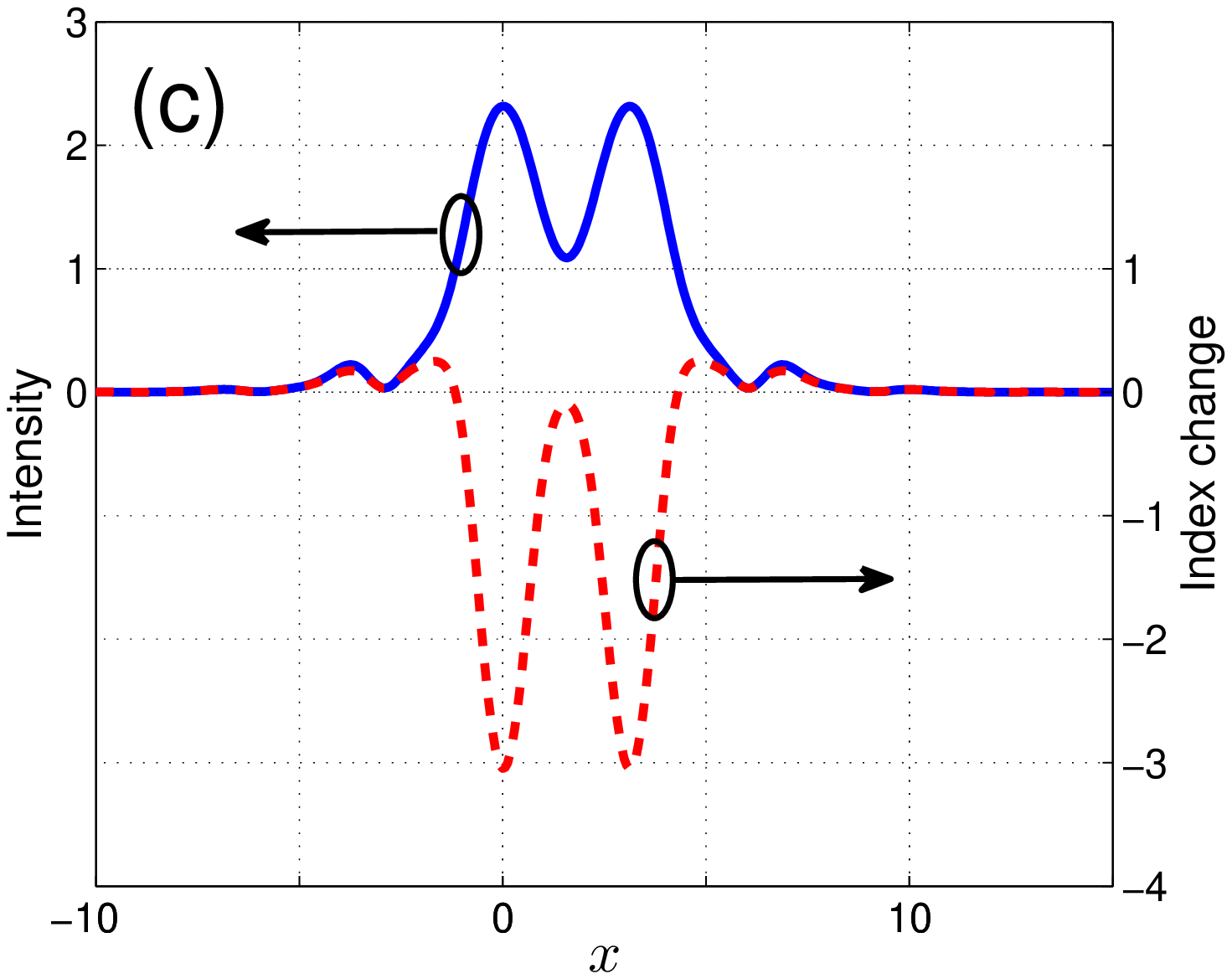}
  \includegraphics[width=0.3\textwidth]{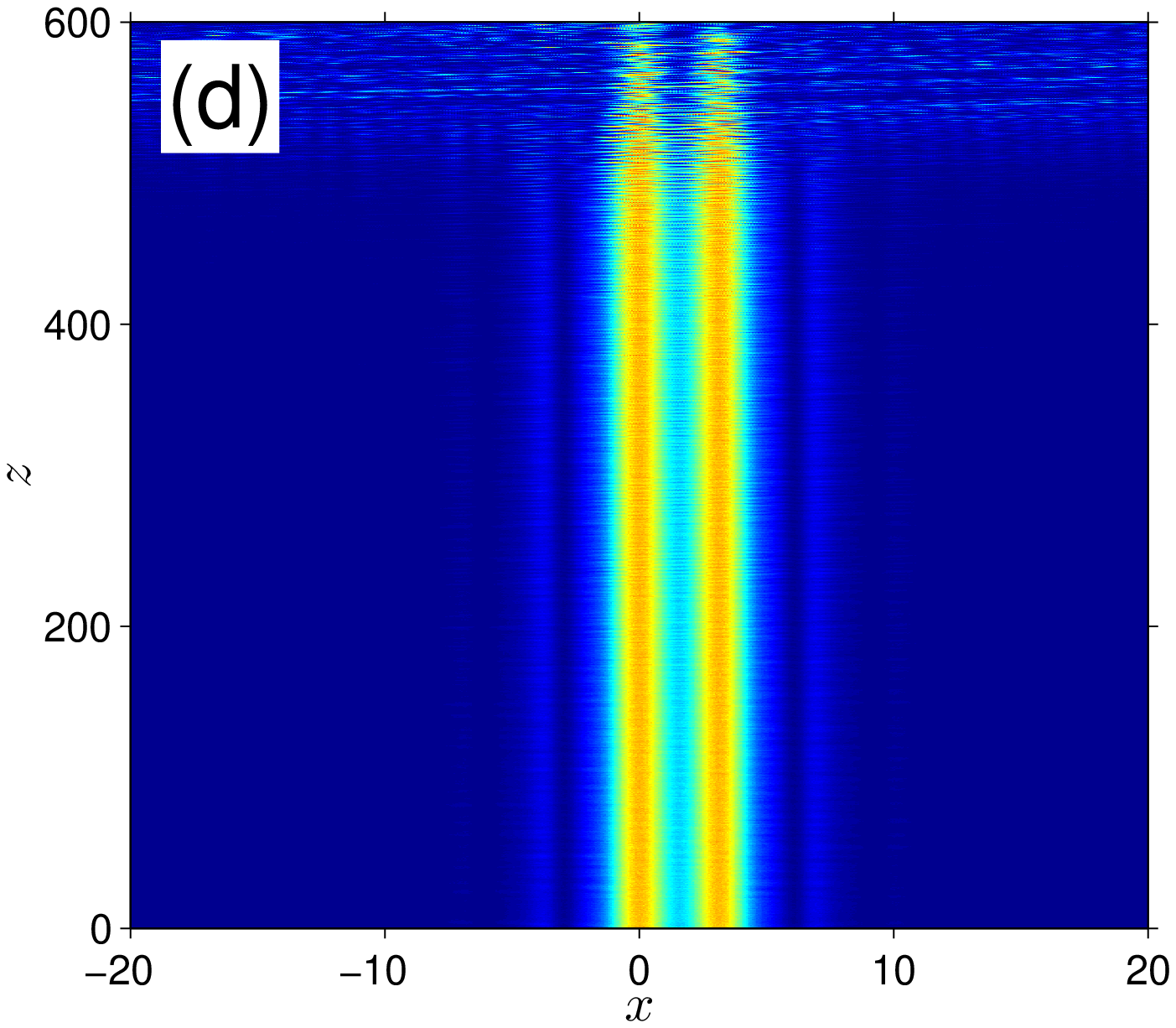}
  \caption{(a) The intensity of the on-site gap soliton (blue curve) and the index change induced by the soliton (red dashed curve);
  (b) Propagation of the soliton shown in (a) with a small perturbation;
  (c) The intensity of the off-site gap soliton (blue curve) and the index change induced by the soliton (red dashed curve);
  (d) Propagation of the soliton shown in (c) with a small perturbation.
  The parameters are $W_0=0.2~\mathrm{and}~\beta=1$ for both solitons.}
  \label{pros}
\end{figure}

As did in the previous work\cite{driben_arxiv_2011a,driben_arxiv_2011b},
stability analysis of the solitons in $\mathcal{PT}$ symmetric potentials discussed in this article is also very important.
In Fig.\ref{pros}, we show the propagation dynamics for the two soliton solutions displayed in Fig.\ref{solitons}(b) and (d)
with small perturbations both on the amplitude and phase.
It is clear to see that the beams annihilate abruptly if the propagation distance exceeds a certain value,
which is similar to the properties of a ``soleakon'' reported in 2009\cite{peleg_pra_2009},
but they do not share the same physical mechanism.
One main difference is that the input used by a soleakon is leaky mode which is unbound state,
while the input used in this article is really bound state.
The reason why the solitons are not stable lies in the fact that defocusing nonlinearity does not support stable bright solitons.
As shown in Fig.\ref{pros}(a) and (d), the index changes (red dashed curves) induced by the humps of the gap solitons (blue curves) are negative,
that means the nonlinearity is defocusing.
Anyway, in a long range (e.g., $0 < z < 150$ for the case shown in Fig.\ref{pros}(b) the solitons are relatively stable.
In a physical point of view,
if we take a soliton beam with $\lambda_0=1.0~\mu \mathrm{m}$, $n_0=1.5$, and $x_0=10~\mu \mathrm{m}$ as the incidence,
it can stably propagate about $2 k_0 n_0 x_0^2 z \approx 28~\mathrm{cm}$ with a diffraction length $L_\mathrm{diff} \approx 1~\mathrm{mm}$.
In comparison with the on-site gap solitons as shown in Fig.\ref{pros}(b),
the off-site gap solitons as shown in Fig.\ref{pros}(d) are much robust.

\begin{figure}[htbp]
\centering
  \includegraphics[width=0.5\textwidth]{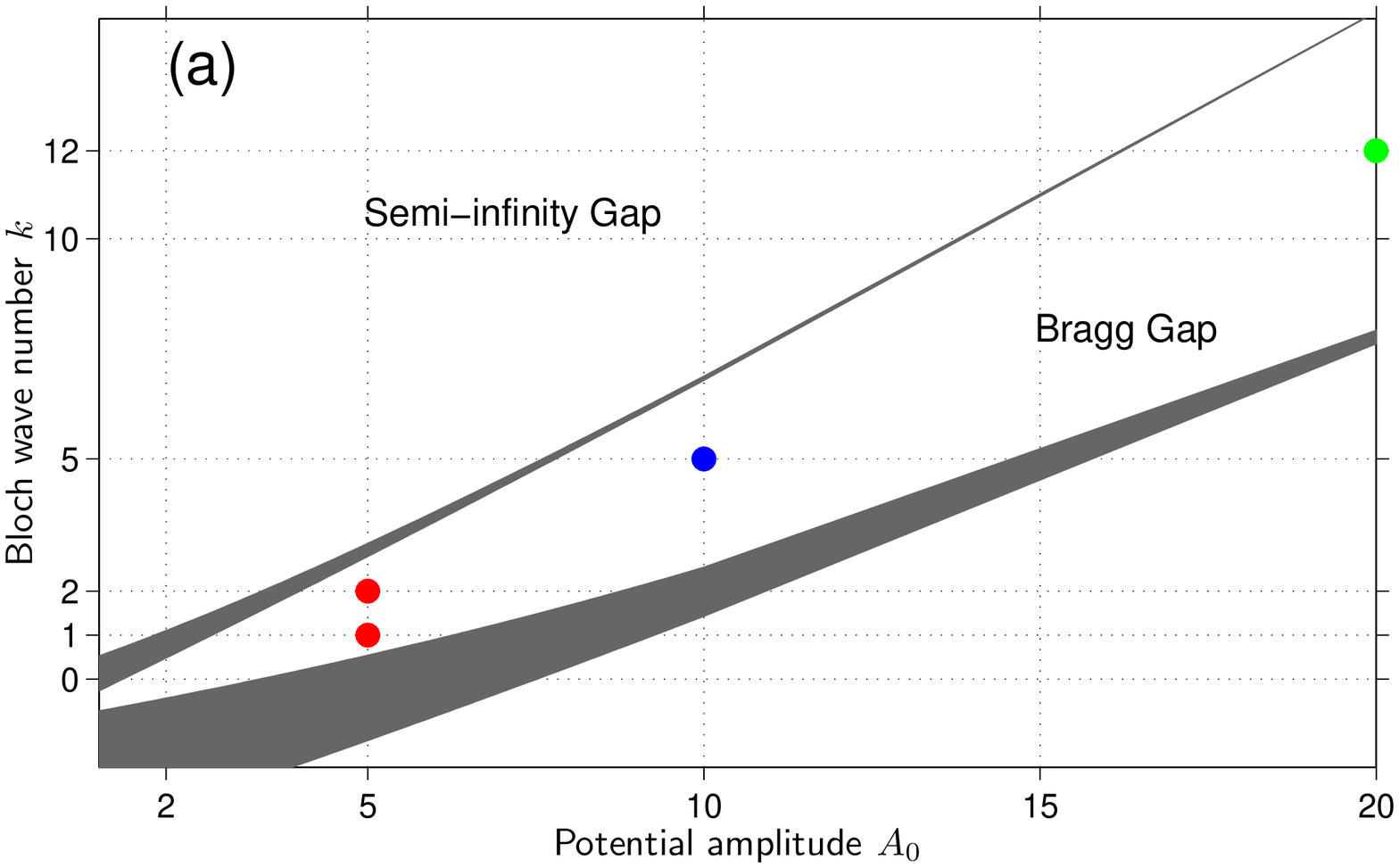} \\
  \includegraphics[width=0.3\textwidth]{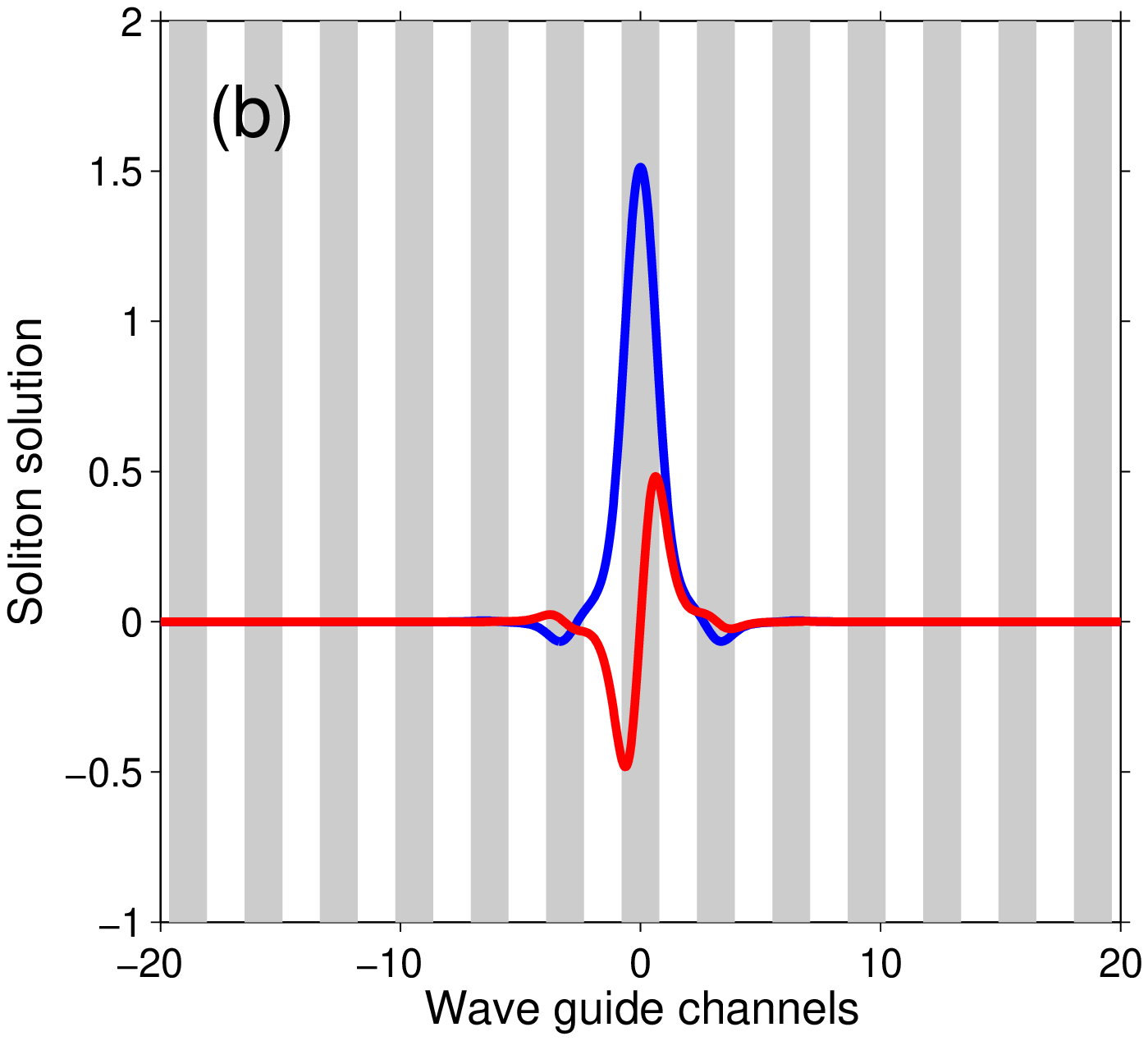}
  \includegraphics[width=0.3\textwidth]{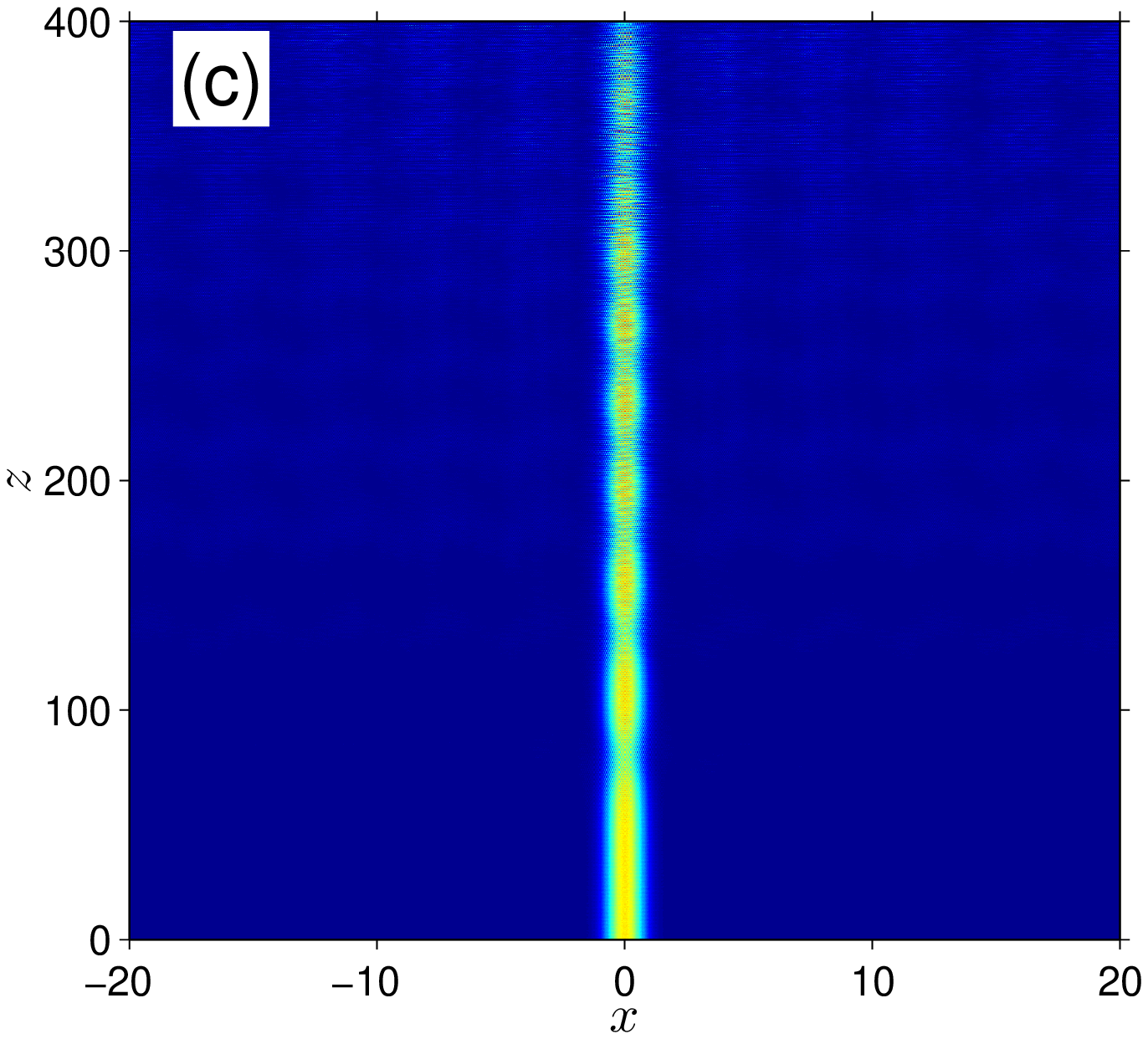}\\
  \includegraphics[width=0.3\textwidth]{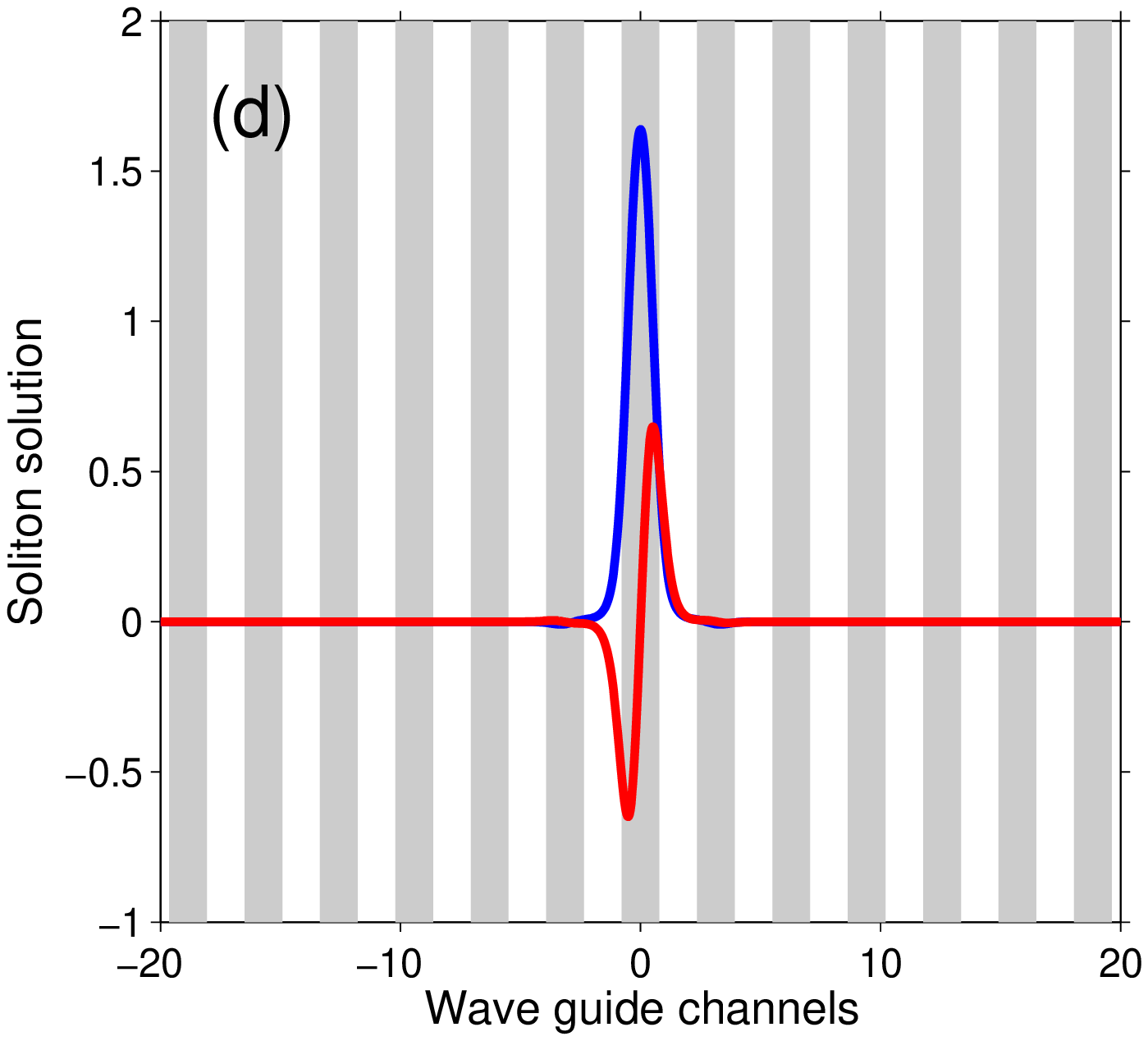}
  \includegraphics[width=0.3\textwidth]{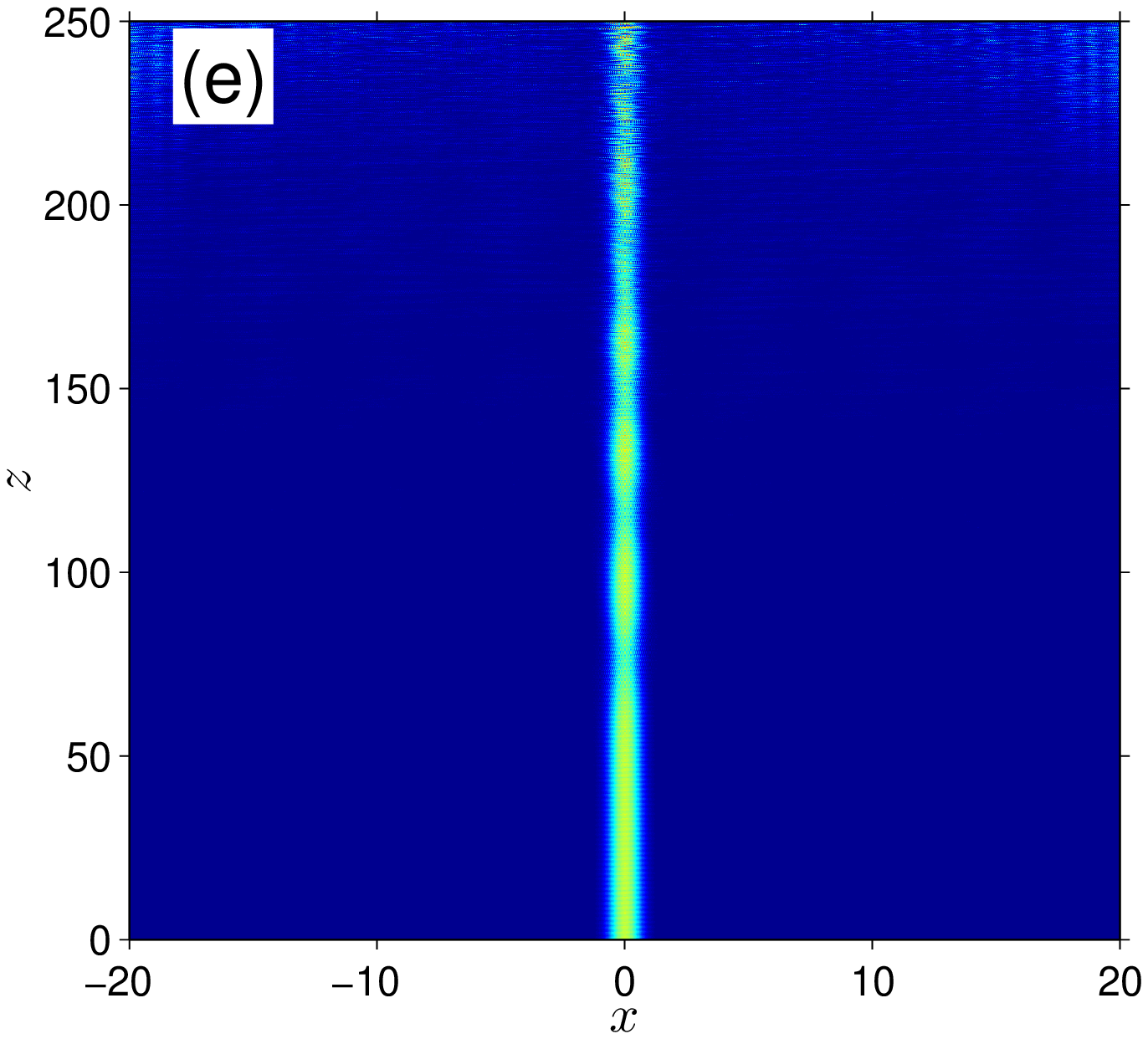}
  \caption{(a) Bragg gap diagram changing with $A_0$ (the parameter $W_0=0.2$),
  in which the semi-infinity gap and Bragg gap regions are marked.
  (b-c) The solitary solution (left panel) and its propagation (right panel) corresponds to the blue dot.
  (d-e) The solitary solution (left panel) and its propagation (right panel) corresponds to the green dot.}
  \label{deeper_potential}
\end{figure}

Naturally, people may ask can the Bragg gap solitons in the complex potentials with higher amplitude $A_0$ be stable during propagation?
To answer this question, we first display the Bragg gap diagram changing with $A_0$ as shown in Fig.\ref{deeper_potential}(a),
in which the two red dots correspond to the unstable solitons we discussed in Fig.\ref{pros} ($A_0=5$).
And then, we discuss another two cases at $A_0=10$ (blue dot) and $A_0=20$ (green dot),
as shown by Figs.\ref{deeper_potential}(b-c) and Figs.\ref{deeper_potential}(d-e), respectively.
According to the propagations, we can conclude that the answer to the question is negative,
for after a relatively stable and breath-like propagation,
the Bragg gap solitons annihilate eventually.

\section{Conclusion}
In conclusion, the existence of Bragg gap solitons in periodic $\mathcal{PT}$ symmetric potentials with
CQ nonlinearity is demonstrated numerically.
Even though the gap solitons we found are not stable, for they will annihilate during propagation
because of defocusing nonlinearity they induced,
they can still stably propagate over a long distance.
Contrast of the two propagations of on-site solitons and off-site solitons,
we find that the latter ones are more robust.

\section*{Acknowledgement}
The author (Yiqi Zhang) thanks the reviewers' illuminating comments
and several friends' nice suggestions (on how to execute numerical simulations more accurately)
to improve the article.


\bibliographystyle{elsarticle-num}
\bibliography{refs}







\end{document}